\documentclass[aps,amsmath,amssymb,pre,twocolumn,showpacs,superscriptaddress]{revtex4-1}
\usepackage{graphicx}
\usepackage{lineno}
\usepackage{natbib}
\usepackage{epstopdf}
\usepackage{amssymb}
\usepackage{url}
\usepackage{amsmath}
\usepackage{color}
\usepackage[caption=false]{subfig}
\usepackage{amssymb}
\usepackage{amsmath}
\usepackage{commath}
\usepackage{graphicx,bm}
\usepackage{verbatim}
\numberwithin{equation}{section}

\begin{document}

\title{Information measure for financial time series:\\ quantifying short-term market heterogeneity}
\author{Linda Ponta}%
\affiliation{Universit\'a Cattaneo LIUC, Castellanza, Italy}
\author{Anna Carbone}
\affiliation{Politecnico di Torino,  corso Duca degli Abruzzi 24, 10129 Torino, Italy}

\begin{abstract}
A  well-interpretable measure of information  has been recently proposed based on  a partition obtained by intersecting  a random sequence with its moving average. The partition yields disjoint sets of the sequence, which are then ranked according to their size to form a probability distribution function and finally fed in the  expression of the Shannon entropy.  In this work, such entropy measure is implemented on the time series of prices and volatilities of six financial markets. The analysis has been performed, on tick-by-tick data sampled every minute for six years of data from 1999 to 2004, for a broad range  of  moving average windows and volatility horizons. The study shows that the entropy of the volatility series depends on the individual market, while  the entropy  of the price series  is practically a market-invariant for the six markets. Finally, a cumulative information measure - the  `Market Heterogeneity Index'- is derived from the integral of the proposed entropy measure. The values of the Market Heterogeneity Index are discussed as possible tools for optimal portfolio construction and compared with those obtained by using  the Sharpe ratio a traditional risk diversity measure.
\end{abstract}

\maketitle

\section{Introduction}

Several connections between economics and statistical thermodynamics have been suggested over the years. Marginal utility and disutility  have been related respectively to
force, energy  and work\cite{fisher1892mathematical}.   Differential
pressure and volume in an ideal gas have been linked  to price and volume in financial systems \cite{samuelson1972maximum}. Analogies have been put forward between
money utility and entropy \cite{lisman1949econometrics} and  between temperature and  velocity of circulation of money \cite{pikler1955utility}.
As a result of the growing diversification and globalization of economy, applications of  entropy concepts to finance and economics are receiving renewed attention as instruments to monitor and quantify market diversity  \cite{chen2015unity,le2006evaluating,ponta2011information,zhou2013applications,ormos2014entropy,yu2014diversified,sheraz2015entropy,pola2016entropy,contreras2017construction,bekiros2017information,chen2017study}.
Heterogeneity of private and institutional investments, might result at a microscopic level,  among other effects, in imperfect and asymmetric flow of information,  ultimately undermining the theory of efficient market  with its assumption of a homogeneous random process underlying the stock prices dynamics \cite{fama1970efficient}.
\par
In the last decades, the interplay of noise and profitability  with specific focus on  volatility investment strategies has therefore gained interest  and is under intense scrutiny   \cite{gunasekarage2001profitability,menkhoff2007obstinate,frommel2016does,smith2016sentiment}. Volatility series exhibit remarkable features suggesting the existence of some amount of `order' out of the seemingly random structure. The degree of `order' is intrinsically linked to the information, embedded in the volatility patterns, whose extraction and quantification might shed light on microscopic phenomena in finance
\cite{zhou2013applications,yu2014diversified,ormos2014entropy,sheraz2015entropy,pola2016entropy,bekiros2017information,contreras2017construction,chen2017study}.
 The Volatility Index (VIX), defined as the near-term volatility conveyed by $S\&P 500$ stock index option prices, has been suggested to monitor investor sentiment. VIX futures and options have been introduced as trading instruments. The Chicago Board Options Exchange (\url{http://www.cboe.com/}) calculates and updates the values of more than 25 indexes designed to measure the expected volatility of different securities.

In this scenario, new tools  for portfolio optimization and asset pricing should be designed on the basis of the expected volatility to evaluate market performances at microscopic level, beyond the figures of risk  provided by more traditional techniques.
\par
In this work, the information measure approach proposed in \cite{carbone2013information,carbone2007scaling,carbone2004analysis} is applied to prices and volatilities of tick-by-tick data,  recorded from 1999 to 2004, of six financial markets. The investigation is performed over a broad range of  volatility and moving average windows. Remarkably, it is found that the entropy of the volatility series takes different values for the different markets as opposed to the entropy of the prices, which is mostly constant and consistent with a homogeneous random walk structure of the time series.
In order to provide a compact visualization of the results and a clear procedure for practically uses of the proposed approach,  a market heterogeneity index (MIX) is introduced, defined on the basis of the integral of the entropy functional. MIX estimates are provided for prices and volatilities of the six financial markets, over different volatility horizons and moving average windows. The values of the index  are finally compared with the results obtained by using the Sharpe ratio, a traditional estimate of portfolio risk.

\section{Entropy}
The definition of entropy, adopted in the framework of this work \cite{carbone2013information,carbone2007scaling,carbone2004analysis}, stem from the idea of Claude Shannon to quantify the ‘expected’  information contained in a message extracted from a sequence $x_t$ \cite{shannon1948mathematical} by using the functional:
\begin{equation}
\label{Shannon}
H[P] = -{S}[P]= -\sum_{j=1}^M p_j\log p_j\hspace{5pt}.
\end{equation}
with $P$ a probability distribution function associated with the time sequence $x_t$. Different approaches have been proposed for the evaluation of the entropy of a random sequence (see  Refs. \cite{grassberger1983characterization,pincus1991approximate,bandt2002permutation,rosso2007distinguishing,crutchfield2012between,marcon2014generalization} for  a few examples of entropy measures). The preliminary but fundamental step is the symbolic representation of the data, through a partition suitable to map the continuous phase-space into disjoint sets. The method commonly adopted for partitioning a sequence  is based on a uniform division in blocks having equal size. Then the entropy is estimated over subsequent partitions corresponding to different block sizes \cite{grassberger1983characterization}. The choice of the optimal partition is not a trivial task, as it is crucial to effectively discriminate between randomness/determinism of the encoded/decoded data  (see Section III of Ref.\cite{daw2003review} for a short review of the partition methods).
\par
\begin{figure}
\includegraphics[width=0.5 \columnwidth]{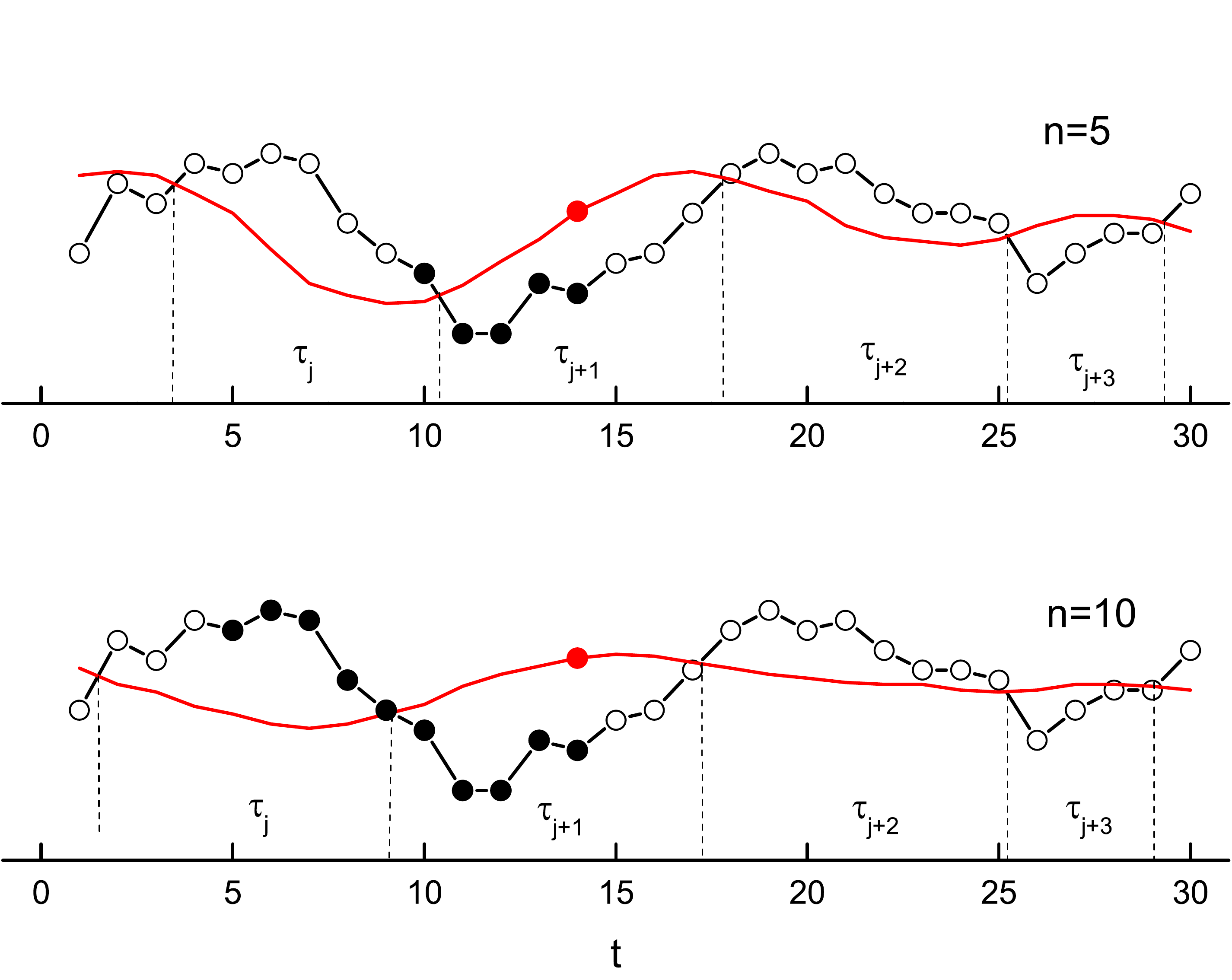}
   \caption{{\textbf{Partition Sketch.}} The illustration shows the partition obtained by the intersection of the time series $\{x_t \}$ and the moving average $\{\widetilde{x}_{t,n}\}$. Top panel is for moving average window $n=5\mathrm{min}$, bottom panel is for moving average window $n=10\mathrm{min}$. The clusters $j, j+1, j+2, j+3$ are shown.}
 \label{Fig:cluster}
\end{figure}

Here, the partition is obtained by taking the intersection of  $\{x_t\}$ with the moving average $\{\widetilde{x}_{t,n}\}$  for different moving average window $n$  \cite{carbone2013information,carbone2007scaling,carbone2004analysis}. For each  window $n$, the subsets $\{x_t: t=s,....,s-n \}$  between two consecutive intersections are ranked according to their size to obtain the probability distribution function $P$. The elements of these subsets are the segments between consecutive intersections and  have been named  \emph{clusters} (see the illustration shown in Fig. \ref{Fig:cluster}).
The present  approach directly yields either power-law or exponential distributed blocks (clusters), thus enabling us to separate the sets of inherently correlated/uncorrelated blocks along the sequence. Moreover, the clusters are exactly defined as the portions of the series between death/golden crosses according to the technical trading rules. Therefore, the information content has a straightforward connection with the trader's view of the price and volatility series.
\par
For the sake of clarity, the main relationships relevant to the investigation carried in this paper will be shortly recalled in the next paragraphs.
\par
Consider the time series  $\{x_t \}$ of length $N$ and the moving average $\{\widetilde{x}_{t,n}\}$   of length $N-n$  with $n$ the moving average window.
The function $\{\widetilde{x}_{t,n}\}$ generates, for each $n$, a partition  $\{\cal{C}\}$  of non-overlapping clusters between two consecutive intersections of $\{x_t \}$ and
$\{\widetilde{x}_{t,n}\}$. Each cluster $j$ has  duration:

\begin{equation}
\label{l} \tau_j\equiv  \|t_{j}-t_{j-1}\|
\end{equation}
\noindent
where the instance $t_{j-1}$ and $t_j$ refer to two subsequent  intersections as shown in Fig.\ref{Fig:cluster}.

\par
 The probability distribution function  $P(\tau,n)$  can be obtained by ranking the number of clusters ${\mathcal N}(\tau_1,n),{\mathcal N}(\tau_2,n), ..., {\mathcal N}(\tau_j,n)$ according to their length $\tau_1, \tau_2,..., \tau_j$ for each $n$.
\par
For a fractional Brownian motion,  a stationary sequence of self-affine clusters $\cal{C}$ is generated with  probability distribution function varying as \cite{carbone2013information}:

\begin{equation}
\label{Pl} P(\tau,n)\sim\tau^{-\alpha} {\mathcal F}\left({\tau},{n}\right) \hspace{5pt},
\end{equation}
with the factor ${\mathcal F}\left({\tau},{n}\right)$ taking the form $ \exp({-\tau}/{n})$,  to account for the finite size effects when $\tau\gg n$, resulting in  the  drop-off of the power-law and the onset of the exponential decay.
\par
By using Eq.~(\ref{Pl}), Eq.~(\ref{Shannon})  writes (the details of the derivation can be found in \cite{carbone2013information,carbone2004analysis}):
\begin{equation}
\label{lentropy2}
S(\tau,n)=S_0+\log\tau^{\alpha}+{\tau\over n}\hspace{5pt},
\end{equation}
where $S_0$ is a constant, $\log\tau^{\alpha}$ and $\tau/ n$  are related respectively to the terms $\tau^{-{\alpha}}$ and ${\mathcal F}(\tau,n)$.
\par
The constant $S_0$ in Eq.~(\ref{lentropy2}) can be evaluated as follows: in the limit $n\sim\tau\rightarrow1$,  $S_0\rightarrow-1$   consistently with the minimum value of the entropy $S(\tau,n)\rightarrow0$ that corresponds to a fully ordered (deterministic) set of clusters with same duration $\tau=1$. The maximum value of  the entropy $S(\tau,n)=\log N^{\alpha}$ is then obtained when $n\sim\tau\rightarrow N$ with $N$  the maximum length of the sequence. This condition corresponds to the maximum randomness (minimum information) carried by the sequence,  when a single cluster is obtained coinciding with the whole series.
\par
Since the exponent $\alpha$ is equal to the fractal dimension $D=2-H$  with $H$ the Hurst exponent of the time series, the term $\log\tau^D$ in Eq.~(\ref{lentropy2}) can be interpreted  as a generalized form of the Boltzmann entropy $S=\log\Omega$, where $\Omega = \tau^D$ can be thought of the volume occupied by the fractional random walker.
\par
The term $\tau/n$ in Eq.~(\ref{lentropy2})
represents  an excess entropy (excess noise) added to the intrinsic entropy term $\log\tau^D$ by the partition process. It depends on $n$ and is related to the finite size effect discussed above. This issue has been discussed in \cite{carbone2013information}, while the general issue of the finite size effects on entropy measure has been discussed  in Refs. \cite{herzel1994finite,grassberger1988finite,stammler2016correcting,levina2017subsampling}
\par
 The method and Eqs.~(\ref{Shannon} - \ref{lentropy2}) have been applied for estimating  the probability distribution function $P(\ell,n)$  and the  entropy $S(\ell,n)$ of the 24 nucleotide sequences of the human chromosomes in  \cite{carbone2013information}. It is worth noting that the characteristic sizes $\ell$  and $\tau$ (respectively cluster length and duration) are equivalent from a statistical point of view. However, from the meaning viewpoint, the cluster duration $\tau$ is more suitable than the length $\ell$ when dealing with time series $\{x_t \}$. Furthermore, the duration  $\tau$ is particularly relevant  to financial market series,  due to its straightforward connection to the duration of the investment horizon via the  volatility window $T$ and the moving average window $n$, entering  as parameters in the above calculation.
\par
Before entering the details of the application of the method to the six financial indexes, it is worthy to clarify the implication  of partitioning a random sequence by either using boxes having equal size or by using the clusters obtained through the intersection with the moving average.
For same size boxes, the excess noise term ${\tau / n}$  vanishes, thus the entropy reduces to the logarithmic term (see Eq. (8) in Ref. ~\cite{grassberger1983characterization}).  The logarithmic term corresponds to the intrinsic entropy of an ideal fractional random walk of dimension $D=2-H$. When a moving average partition is used, an excess entropy term  ${\tau / n}$  emerges accounting for the additional heterogeneity introduced by the random partitioning process (operated by the moving average partition). This is not a detail especially for financial applications as it represents the disorder specifically introduced by this trading mechanism.

\section{Results}
\label{SecDataSet}
The information measure described in the previous section is here applied to financial prices and volatilities of six market indices (BOBL, BUND, DAX, Euro Currency, Euro Stoxx and FIB30).
For each index, the data set includes tick-by-tick prices $p_t$ sampled every minute from 4 January 1999 to 24 March 2004. The  length of the series is set for all the six markets equal to 517041, determined on the basis of the length of the shortest series which is the Eurocurrency.  The six  time series are shown in Fig.~\ref{Fig:PRICE}. Further details are provided in the Supplementary Material.
\par
Among other reasons, the main motivation for selecting  the six markets mentioned above is the fact that they operate in a very similar socio-economic context, being all traded within the EU zone. This rules out that the diversity featured by the entropy measure might stem from exogenous drives  rather than from the intrinsic dynamics of the market.
\par
In this work, the continuously compounded return:
\begin{equation}
 r_t = p_t - p_{t-h} \hspace{5pt} ,
\end{equation}
and the log-return:
\begin{equation}
r_t  = \log p_t - \log p_{t-h} \hspace{5pt},
\end{equation}
have been considered, with $ 0<h<t<N $ and $N$ the maximum length of the time series.
\par
The volatility $v_{t,T}$ has been taken as:
\begin{equation}
v_{t,T}=\sqrt{\frac{\sum_{t=1}^{T}{(r_t-\mu_T)^2}}{T- 1} } \hspace*{5 pt},
\label{volatility}
\end{equation}
where $r_t$ are the above defined  return, $T$  the volatility window, $\mu_T=\sum_{t=1}^{T}{r_t}/(T-1)$ the average value of the return taken over the window $T$. It is well known that the volatility definition is not univocal, however  this work has been focused on Eq.~(\ref{volatility}).

\par
\begin{figure}
\includegraphics[width=\columnwidth]{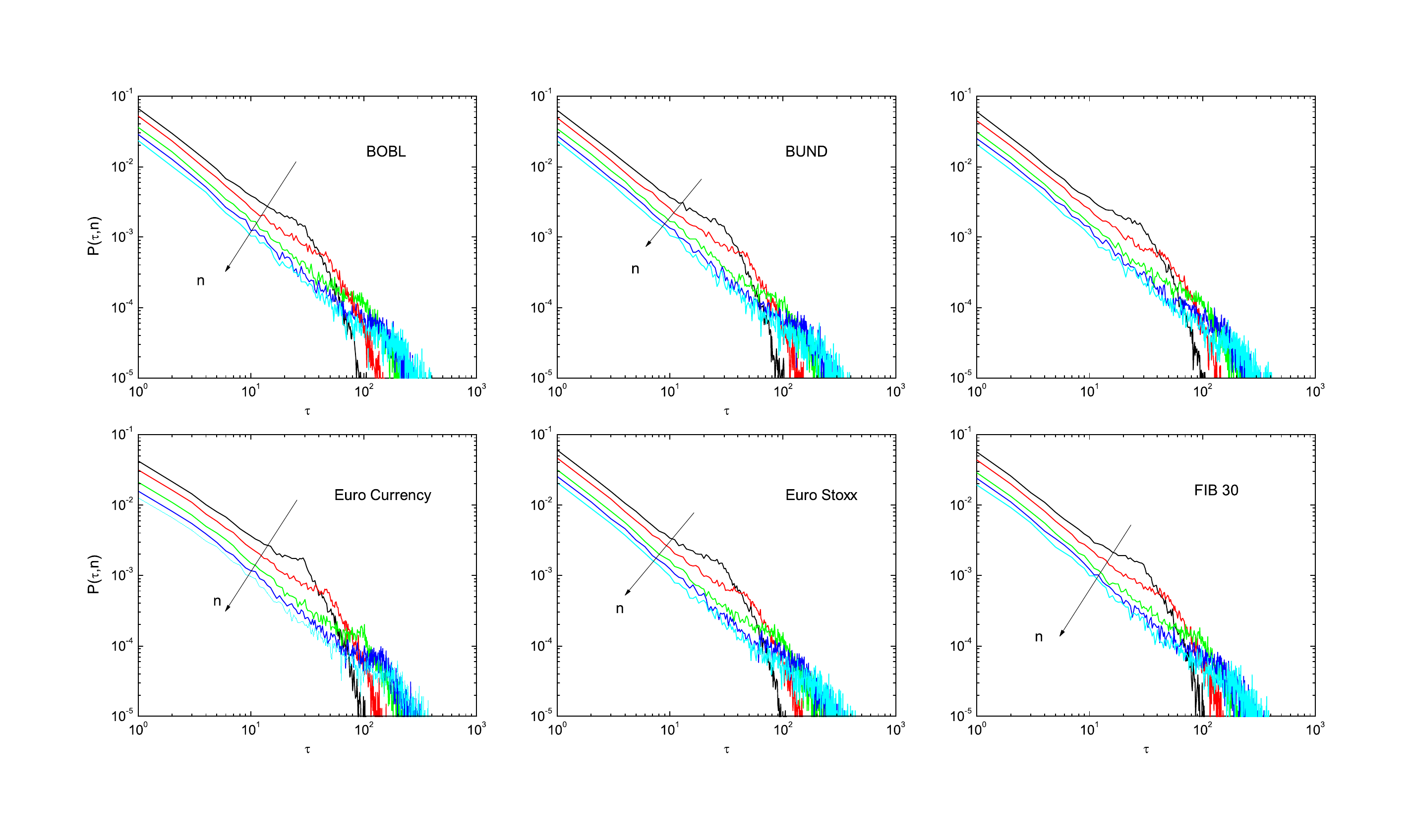}
 \caption{\textbf{Probability distribution function for the price series.} Probability distribution function ${P}(\tau,n)$ for the prices series of the BOBL, BUND, DAX, Euro Currency, Euro Stoxx and FIB30 markets. The different plots refer to different values of the moving average window $n$ (namely $n=30\hspace{2pt}\mathrm{m in}$, $n=50\hspace{2pt}\mathrm{min}$, $n=100\hspace{2pt}\mathrm{min}$, $n=150\hspace{2pt}\mathrm{min}$ and $n=200\hspace{2pt}\mathrm{min}$).}
 \label{Fig:PDF}
\end{figure}

Next, the series of the prices of the  six markets, shown in Fig.~\ref{Fig:PRICE}, are used to calculate the probability distribution functions $P(\tau,n)$ and the entropy ${S}(\tau,n)$ according to the procedure described in the previous section.  The probability $P(\tau,n)$  for the six price series is shown in Fig.~\ref{Fig:PDF}. For the clarity of visualization,  only the curves corresponding to moving average windows  $n=30\hspace{2pt}\mathrm{min}$, $n=50\hspace{2pt}\mathrm{min}$, $n=100\hspace{2pt}\mathrm{min}$, $n=150\hspace{2pt}\mathrm{min}$, $n=200 \hspace{2pt} \mathrm{min}$ are plotted, though the analysis has been performed for  $n$ ranging  between $5\hspace{2pt}\mathrm{min}$ and $1500\hspace{2pt}\mathrm{min}$.  The curves exhibit a  power law behaviour  for $\tau <n$, and an exponential decay,  for $\tau >n$, respectively. The exponents agree with the theoretical prediction $\alpha=2-H$  expected for fractional Brownian motions. The curves shown in  Fig.~\ref{Fig:PDF} are thus consistent with Eq.~(\ref{Pl}) over the whole range of investigated values.
\par
\begin{figure}
\includegraphics[width=\columnwidth]{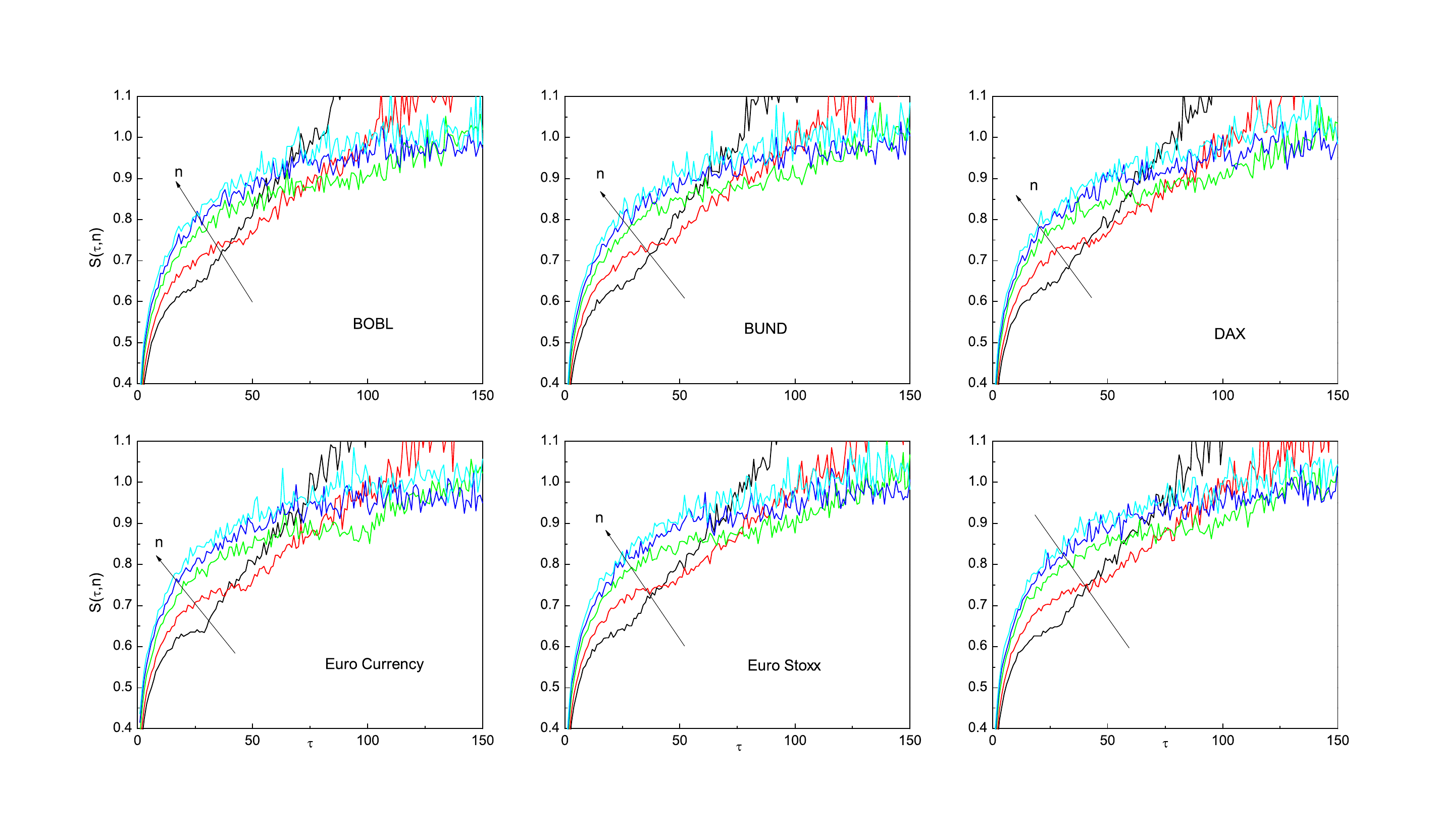}
\caption{{\bf Entropy for the price series.} Entropy ${S}(\tau,n)$  for the time series of the prices of the BOBL, BUND, DAX, Euro Currency, Euro Stoxx and FIB30  markets. The different plots refer to different values of the moving average window $n$ (namely $n=30\mathrm{min}$, $n=50\mathrm{min}$, $n=100\mathrm{min}$, $n=150\mathrm{min}$ and $n=200\mathrm{min}$).}
 \label{Fig:ENTROPYP}
\end{figure}

Fig.~\ref{Fig:ENTROPYP} shows the entropy $S(\tau ,n)$ calculated by using the probability distribution functions plotted in Fig.~\ref{Fig:PDF}. The curves correspond to moving average values  $n=30\hspace{2pt}\mathrm{min}$,  $n=50\hspace{2pt}\mathrm{min}$,  $n=100\hspace{2pt}\mathrm{min}$,  $n=150\hspace{2pt}\mathrm{min}$,  $n=200\hspace{2pt}\mathrm{min}$. The range of investigated moving average windows is much broader (from $n=5\hspace{2pt}\mathrm{min}$ to $n=1500\hspace{2pt}\mathrm{min}$) but it is  not shown for visualization clarity. The behavior of the curves is quite carefully reproduced by  Eq.~(\ref{lentropy2}): the first part  increases as a logarithmic function, then for $\tau \approx n$ a sharp increase is observed corresponding to the onset of the linear term ${\tau /n}$.
One can note that ${S}(\tau,n)$
 is $n$-invariant for small values of $\tau$, while its slope decreases as $1/n$  at larger $\tau$, as expected according to Eq.~(\ref{lentropy2}), meaning that clusters with duration $\tau > n $
are not power-law correlated, due to the finite-size
effects introduced by the partition with window $n$. Hence, they are characterized
by a value of the entropy exceeding the curve $\log \tau^D$, which corresponds to power-law correlated clusters.
 It is worthy to remark that clusters with same duration $\tau$
  can be generated by different values of the moving average window $n$.
For constant $\tau$, larger entropy values are obtained as $n$ increases.
\par
Next, the probability distribution function $P(\tau,n)$ and the entropy $S(\tau,n)$ have been calculated for a large set of volatility series.   $P(\tau,n)$ and $S(\tau ,n)$   have been calculated  for (i) linear return   and (ii)  logarithmic return with volatility window $T$ ranging from half a trading day to  20 trading days.  In particular $T=330\hspace{2pt}\mathrm{min}$, $T=660 \hspace{2pt}\mathrm{min}$, $T=1320\hspace{2pt}\mathrm{min}$, $T=1980\hspace{2pt}\mathrm{min}$, $T=2640\hspace{2pt}\mathrm{min}$, $T=3300\hspace{2pt}\mathrm{min}$, $T=3960\hspace{2pt}\mathrm{min}$, $T=4620\mathrm{min}$, $T=5280\hspace{2pt}\mathrm{min}$, $T=5940\hspace{2pt}\mathrm{min}$, $T=6600\hspace{2pt}\mathrm{min}$, $T=13200\hspace{2pt}\mathrm{min}$ have been analysed, that correspond to half a business day, one to ten business days and one business month, respectively.
\par
\begin{figure}
\includegraphics[width=\columnwidth]{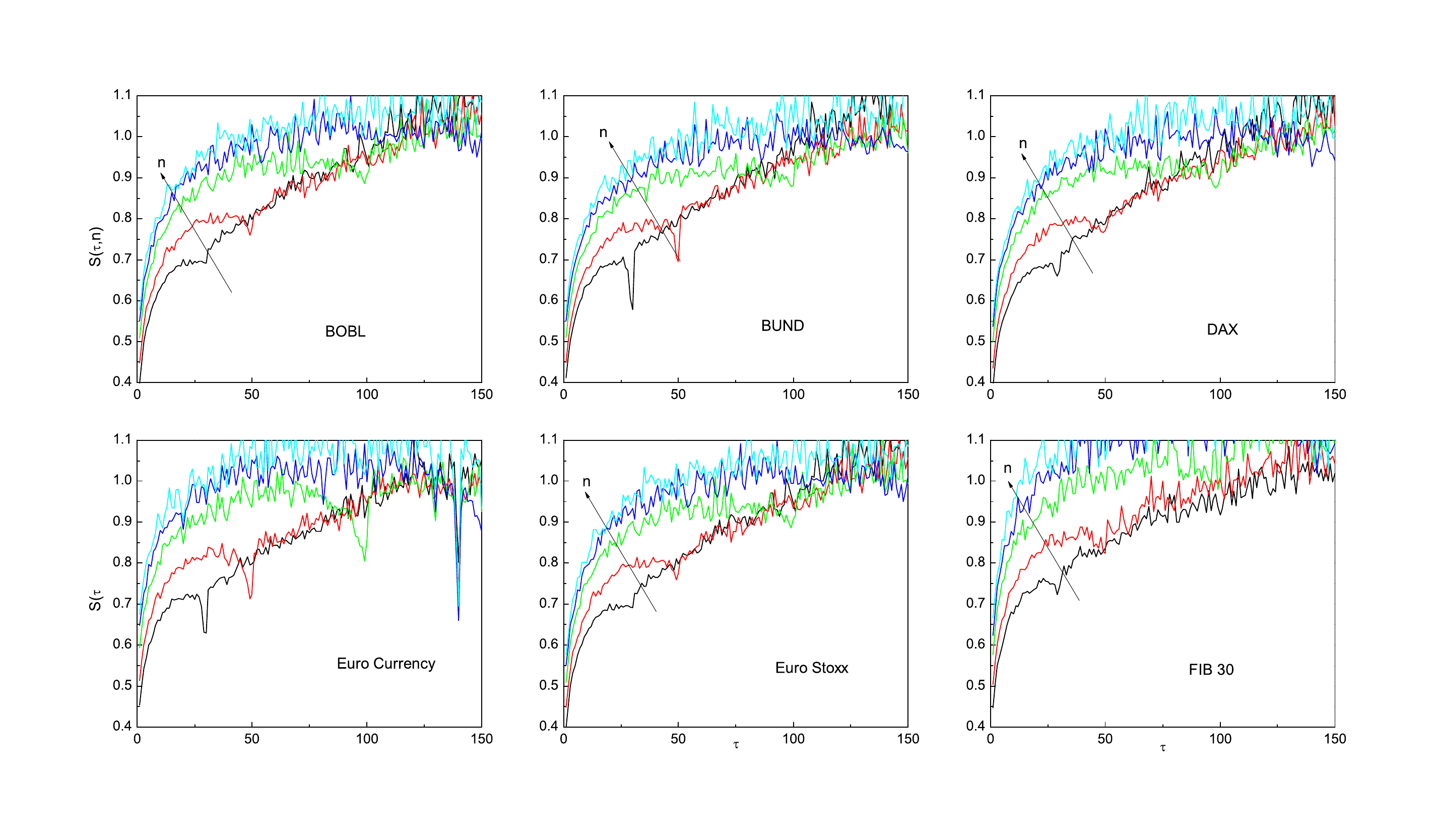}
\caption{{\bf Entropy for the volatility series.} The function  ${S}(\tau,n)$  is shown  for the volatilities of the log return time series of the BOBL, BUND, DAX, Euro Currency, Euro Stoxx and FIB30 markets. The volatility window $T=660\hspace{2pt}\mathrm{min}$  for all the six graphs. The different plots refer to different values of the moving average window $n$ (namely $n=30\hspace{2pt}\mathrm{min}$, $n=50\hspace{2pt}\mathrm{min}$, $n=100\hspace{2pt}\mathrm{min}$, $n=150\hspace{2pt}\mathrm{min}$ and $n=200\hspace{2pt}\mathrm{min}$).}
 \label{Fig:ENTROPYV}
\end{figure}

 The entropy ${S}(\tau,n)$  for the volatility of  logreturn with $T=660\hspace{2pt}\mathrm{min}$ is shown in Fig.~\ref{Fig:ENTROPYV} for the six markets. By comparing the results shown in Fig.~\ref{Fig:ENTROPYV} and in Fig.~\ref{Fig:ENTROPYP},  one can note a general increase of the entropy curves of the  volatilities compared to those of the prices. Such an increase has been systematically found in all the series analysed, both for linear and logarithmic return with diverse volatilities windows $T$.
\par
The sharp drop of the entropy is related to a deviation of the probability distribution function of the real time series from the behavior predicted on the basis of the Eq. (3) deduced  for an ideal fractional Brownian motion. The deviation is related to the reduced randomness and data redundancy in the series. These issues are in particular enhanced when the data are fed in the volatility relationship (Eq. (7)) which, as a variance,  enhances the variability compared to that of the prices.

\begin{figure}
\includegraphics[width=\columnwidth]{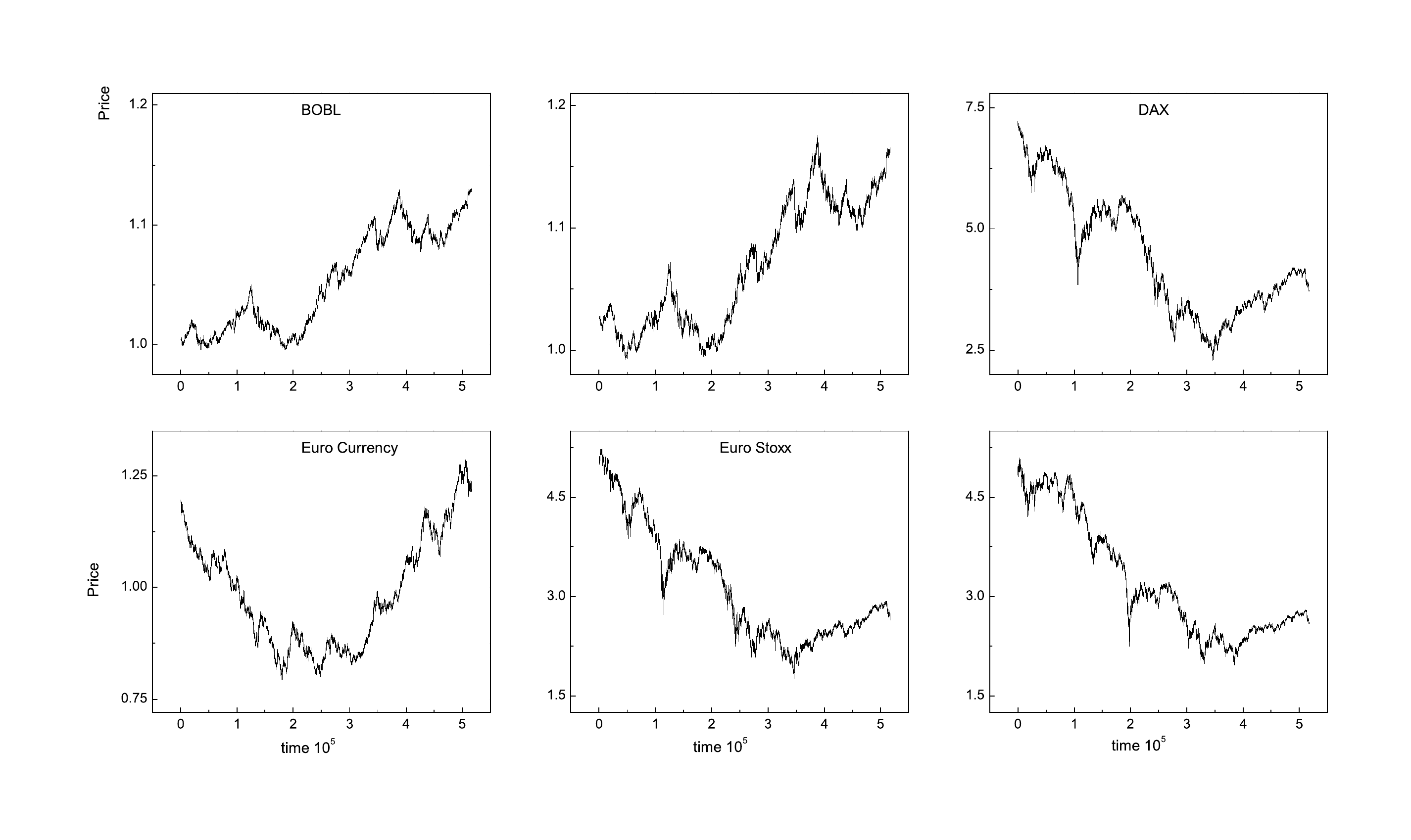}
\caption{\textbf{Prices Series.}  Prices data for the  BOBL, BUND, DAX, Euro Currency, Euro Stoxx and FIB30  indexes. The tick-by-tick data series are sampled every minute from 4 January 1999 to 24  March 2004. The total length of the series is 517041. Further details concerning the six marketplaces are provided in the Supplementary Material \ref{S1}.}
 \label{Fig:PRICE}
\end{figure}

\begin{figure}
\includegraphics[width=\columnwidth]{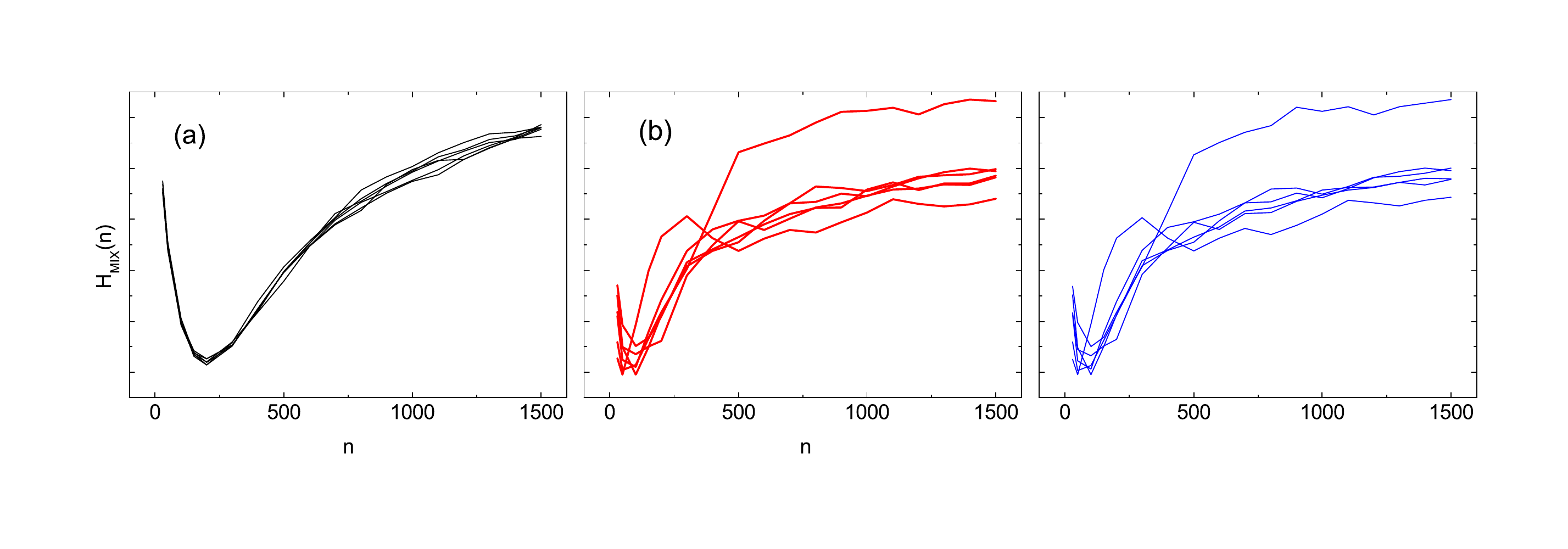}
\caption{{\bf } Plot of the function $H_{MIX}(n)$  against $n$ calculated according to Eq.~(\ref{HMIX}) for the prices (a), the volatilities with linear return  (b) and  with log return  (c) for the six indexes. Consistently with the results shown in Fig.~(\ref{Fig:ENTROPYP}) the plots exhibit an increasing behaviour with an inversion at small values of $n$. A strong variability is observed for the volatilities (graphs (b) and (c)) as opposed to the prices series (a).}
 \label{Fig:HMIXn}
\end{figure}

\begin{figure}
\includegraphics[width=\columnwidth]{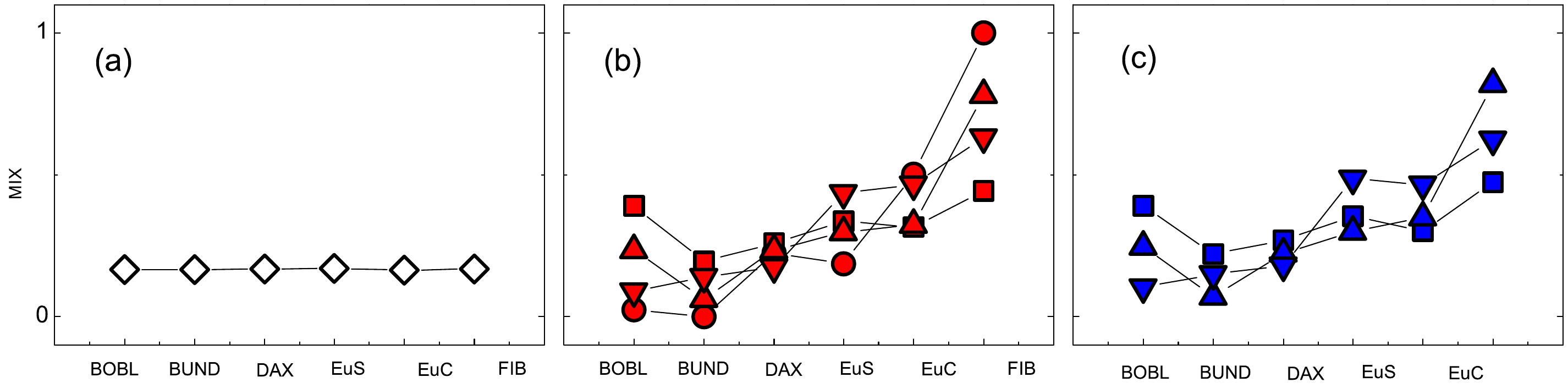}
\caption{Market Heterogeneity Index calculated according to Eq.~(\ref{MIX}) for the  the six indexes. The volatility window is  $T=1320\hspace{2pt}\mathrm{min}$ (circles), $T=1980\hspace{2pt}\mathrm{min}$ (squares),  $T=2640\hspace{2pt}\mathrm{min}$ (up triangles) and  $T=3300\hspace{2pt}\mathrm{min}$ (down triangles). A strong variability of the MIX is observed  for the volatility series (b) and (c), as opposed to the price series exhibiting a constant value of the index (a). The values of the MIX has been rescaled between 0 and 1.}
\label{Fig:MIX}
\end{figure}

\begin{figure}
\includegraphics[width=0.4\columnwidth]{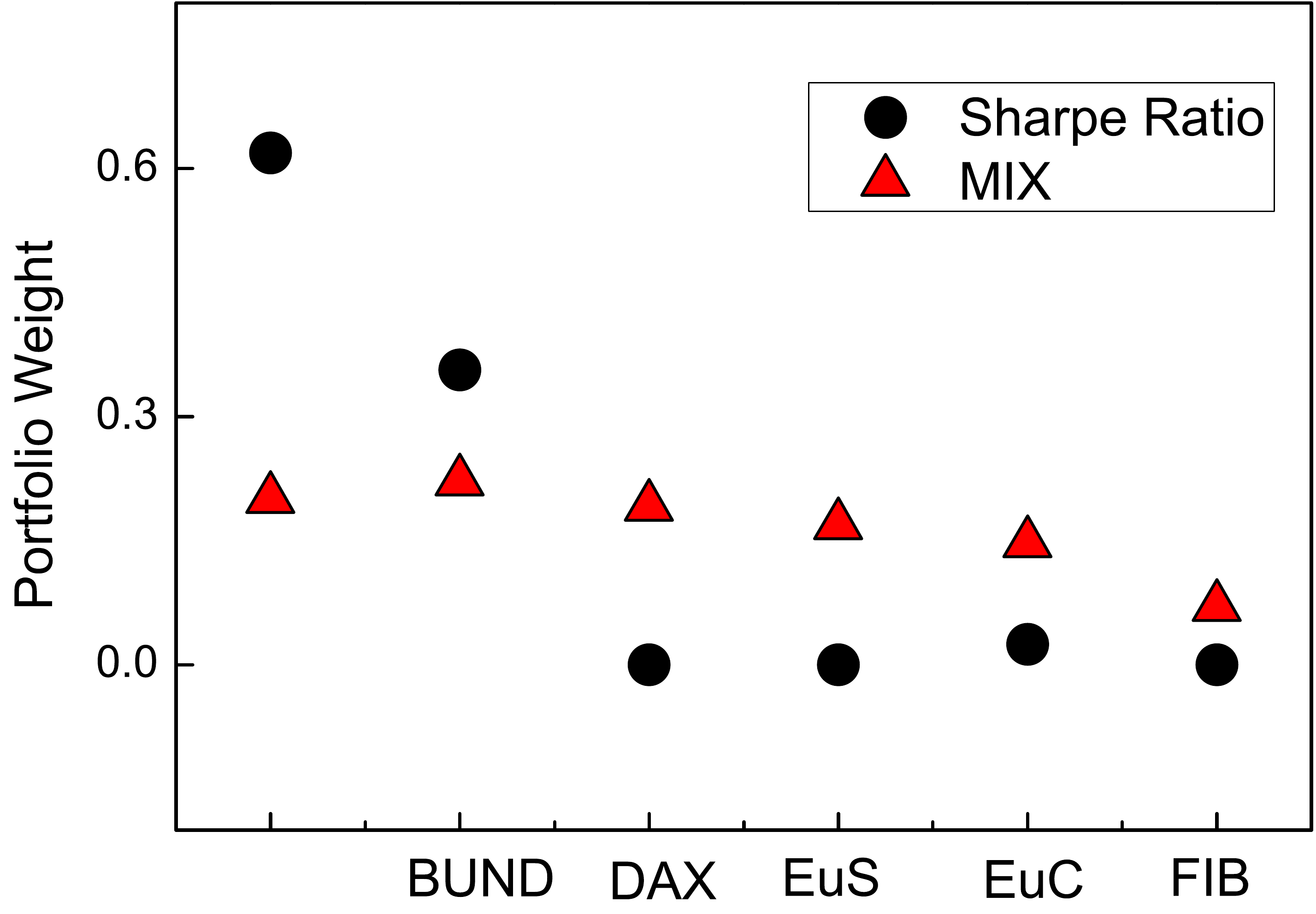}
\caption{Portfolio weights obtained by using the Sharpe ratio and  the Market Heterogeneity Index defined by Eqs.~(\ref{HMIX}) for the six indexes.}
\label{Fig:WEIGHT}
\end{figure}

\par
To better quantify the behaviour  of prices and volatilities of the different markets, we have introduced a cumulative information heterogeneity index defined as follows:
\begin{equation}
H_{MIX}(n)=-\int_0^{\tau_{max}} S(\tau,n)d\tau \hspace*{5 pt}.
\label{HMIX}
\end{equation}
The function $H_{MIX}(n)$ has been plotted in Fig.~\ref{Fig:HMIXn} for the prices (a) and the volatilities with linear  (b) and logarithmic return (c).  The function $H_{MIX}(n)$ is almost constant for the prices of the six markets (a),  conversely it takes different values for the volatilities (b,c).
\par
In order to compare the obtained results with traditional portfolio risk estimator, the quantity ${MIX}$ has been calculated as the integral of the function $H_{MIX}(n)$ over $n$:
\begin{equation}
MIX=\int_{n_{min}}^{n_{max}} H_{MIX}(n) dn \hspace*{5 pt}.
\label{MIX}
\end{equation}
The MIX is a cumulative figure  of heterogeneity  suitable to quantify the reduced information content (excess entropy) and make a comparison  across the different markets. The absolute value of the market heterogeneity index (MIX) has been plotted in Fig.~\ref{Fig:MIX} for the six markets. In particular the data in Fig.~\ref{Fig:MIX} (a) corresponds to integrals under the curves shown in Fig.~\ref{Fig:HMIXn} (a) for the prices. The data of  Fig.~\ref{Fig:MIX} (b) correspond to the integrals of the volatility curves with linear return shown in Fig.~\ref{Fig:HMIXn} (b).  The data of Fig.~\ref{Fig:MIX} (c) correspond to the integrals of the volatility curves  with the logarithmic return shown in Fig.~\ref{Fig:HMIXn} (c).
Different symbols corresponds to the different volatility windows, namely $T=1320\hspace{2pt}\mathrm{min}$ (circles), $T=1980\hspace{2pt}\mathrm{min}$ (squares), $T=2640\hspace{2pt}\mathrm{min}$ (up triangles), $T=3300\hspace{2pt}\mathrm{min}$ (down triangles). Consistenly with the results shown in Fig.~\ref{Fig:HMIXn}, a strong variability of the MIX is observed for the volatility curves (b,c) compared to prices (a).
It is worthy of note that the investment horizon of the current entropy measure is defined by the range of the cluster duration $\tau$  i.e  of the order or smaller than the maximum moving average window $n$. Conversely, the optimization based on the Sharpe ratio refers to the whole duration of the financial data series. Therefore the time scale of the proposed measure  refers to a short term investment horizon (day-by-day trading).
\par
Next we will discuss the issue that the information heterogeneity index evaluated on volatility (with linear or logarithmic return) can be used for constructing the  weights of a portfolio asset.  By using the MIX strategy, the fraction of investments in each market normalized to 1 can be obtained by taking the values $1-MIX$.
These values are shown in  Fig.~\ref{Fig:WEIGHT} (up triangles) for the MIX values shown in  Fig.~\ref{Fig:MIX} (b).
In Fig.~\ref{Fig:WEIGHT}  the weights of an efficient portfolio that maximize the Sharpe Ratio are shown for comparison (circles).  The Sharpe Ratio is a measure for calculating risk-adjusted return. It is the average return earned in excess of the risk-free rate per unit of volatility or total risk:
\begin{equation}
SR=\frac{(\overline{r}_p -r_f)}{\sigma_p}\hspace{5pt} ,
\end{equation}
  where  $\overline{r}_p$ is the expected portfolio return,   $r_f$  the risk free rate  and  $\sigma_p$  the portfolio standard deviation.  Generally, the greater the value of the Sharpe ratio, the more attractive the risk-adjusted return. The values have been obtained by using the Sharp Ratio optimazation tool provided by MATLAB. The portfolio weights obtained via the MIX startegy and the Sharpe ratio are shown in Table 1.  There is a very good correspondence between the weights obtained with the two strategy. It is worth noting that the portfolio that maximizes the Sharpe ratio correspond to a strategy limited to only 3 out of the 6 markets, as opposed to the MIX strategy that has a smoother variation of the weights.

%

\section*{Conclusion}
The implementation of the moving average  algorithm to estimate the Shannon entropy of a long-range correlated sequence has been illustrated to analyse the tick-by-tick data of six markets BOBL, BUND, DAX, Euro Currency, Euro Stoxx and FIB30 from 1999 to 2004. By considering several runs of the algorithm for the different parameters (mainly moving average $n$ and volatility $T$ windows) this study has systematically shown, the entropy measured on the price is practically market invariant, whereas the entropy of the volatilities is market dependent.  A cumulative market heterogeneity index (MIX) has been built and its values compared  with those of a standard risk measure (the Sharpe ratio).  Compared to the Sharpe ratio, the MIX index provide a more accurate and smooth evaluation of the portfolio composition, that might be an advantage for practical applications of the entropy measure.
The novelty  of the present  approach resides in the method used for partitioning the sequence which  allows one to separate the sets of inherently informative/uninformative clusters along the sequence. Moreover, the clusters are exactly defined as the portions of the series between death/golden crosses according to the technical traders rules. Therefore, the information content  quantified by this approach is intimately linked to technical trader's viewpoint on the analysed markets.
The six markets used in this study  have been chosen because they  operate in close  socio-economic contexts within the EU.  The similarity of the external context permits to rule out external causes as the source of the market heterogeneity, while ensuring that the diversity is due  to endogenous processes.  The current work will be extended to evaluate extended portfolio including markets beyond Europe and other volatile assets.

\begin{thebibliography}{10}

\bibitem{fisher1892mathematical}
Fisher I.
\newblock Mathematical investigations in the theory of value and prices.
\newblock Yale University. 1892.

\bibitem{samuelson1972maximum}
Samuelson PA.
\newblock Maximum principles in analytical economics: Nobel Lecture 1970.
\newblock The American Economic Review. 1972;62(3):249--262.

\bibitem{lisman1949econometrics}
Lisman JH.
\newblock Econometrics and thermodynamics: a remark on Davis' theory of
  budgets.
\newblock Econometrica: Journal of the Econometric Society. 1949; p. 59--62.

\bibitem{pikler1955utility}
Pikler AG.
\newblock Utility theories in field physics and mathematical economics (II).
\newblock The British Journal for the Philosophy of Science.
  1955;5(20):303--318.

\bibitem{chen2015unity}
Chen J.
\newblock The unity of science and economics: A new foundation of economic
  theory.
\newblock Springer; 2015.

\bibitem{le2006evaluating}
Le~Gallo J, Dall’Erba S.
\newblock Evaluating the temporal and spatial heterogeneity of the European
  convergence process, 1980--1999.
\newblock Journal of Regional Science. 2006;46(2):269--288.

\bibitem{ponta2011information}
Ponta L, Pastore S, Cincotti S.
\newblock Information-based multi-assets artificial stock market with
  heterogeneous agents.
\newblock Nonlinear Analysis: Real World Applications. 2011;12(2):1235 -- 1242.
\newblock doi:{http://dx.doi.org/10.1016/j.nonrwa.2010.09.018}.

\bibitem{zhou2013applications}
Zhou R, Cai R, Tong G.
\newblock Applications of entropy in finance: A review.
\newblock Entropy. 2013;15(11):4909--4931.

\bibitem{ormos2014entropy}
Ormos M, Zibriczky D.
\newblock Entropy-based financial asset pricing.
\newblock PloS one. 2014;9(12):e115742.

\bibitem{yu2014diversified}
Yu JR, Lee WY, Chiou WJP.
\newblock Diversified portfolios with different entropy measures.
\newblock Applied Mathematics and Computation. 2014;241:47--63.

\bibitem{sheraz2015entropy}
Sheraz M, Dedu S, Preda V.
\newblock Entropy Measures for Assessing Volatile Markets.
\newblock Procedia Economics and Finance. 2015;22:655--662.

\bibitem{pola2016entropy}
Pola G.
\newblock On entropy and portfolio diversification.
\newblock Journal of Asset Management. 2016;17(4):218--228.

\bibitem{contreras2017construction}
Contreras J, Rodr{\'\i}guez YE, Sosa Al.
\newblock Construction of an efficient portfolio of power purchase decisions
  based on risk-diversification tradeoff.
\newblock Energy Economics. 2017;64:286--297.

\bibitem{bekiros2017information}
Bekiros S, Nguyen DK, J~L S, Uddin GS.
\newblock Information diffusion, cluster formation and entropy-based network
  dynamics in equity and commodity markets.
\newblock European Journal of Operational Research. 2017;256(3):945--961.

\bibitem{chen2017study}
Chen X, Tian Y, Zhao R.
\newblock Study of the cross-market effects of Brexit based on the improved
  symbolic transfer entropy GARCH model—An empirical analysis of stock--bond
  correlations.
\newblock PloS one. 2017;12(8):e0183194.


\bibitem{fama1970efficient}
Fama EF.
\newblock Efficient capital markets: A review of theory and empirical work.
\newblock The Journal of Finance. 1970;25(2):383--417.

\bibitem{gunasekarage2001profitability}
Gunasekarage A, Power DM.
\newblock The profitability of moving average trading rules in South Asian
  stock markets.
\newblock Emerging Markets Review. 2001;2(1):17--33.

\bibitem{menkhoff2007obstinate}
Menkhoff L, Taylor MP.
\newblock The obstinate passion of foreign exchange professionals: technical
  analysis.
\newblock Journal of Economic Literature. 2007;45(4):936--972.

\bibitem{frommel2016does}
Fr{\"o}mmel M, Lampaert K.
\newblock Does frequency matter for intraday technical trading?
\newblock Finance Research Letters. 2016;18:177--183.

\bibitem{smith2016sentiment}
Smith DM, Wang N, Wang Y, Zychowicz EJ.
\newblock Sentiment and the effectiveness of technical analysis: Evidence from
  the hedge fund industry.
\newblock Journal of Financial and Quantitative Analysis.
  2016;51(6):1991--2013.



\bibitem{carbone2013information}
Carbone A.
\newblock Information Measure for Long-Range Correlated Sequences: the Case of
  the 24 Human Chromosomes.
\newblock Scientific Reports. 2013;3:2721.

\bibitem{carbone2007scaling}
Carbone A, Stanley HE
\newblock Scaling properties and entropy of long-range correlated time series
\newblock  Physica A: Statistical Mechanics and its Applications 384 (1), 21-24


\bibitem{carbone2004analysis}
Carbone A, Castelli G, Stanley HE.
\newblock Analysis of clusters formed by the moving average of a long-range
  correlated time series.
\newblock Phys Rev E. 2004;69:026105.
\newblock doi:{10.1103/PhysRevE.69.026105}.

\bibitem{shannon1948mathematical}
Shannon CE.
\newblock A mathematical theory of communication, Part I, Part II.
\newblock Bell Syst Tech J. 1948;27:623--656.

\bibitem{grassberger1983characterization}
Grassberger P, Procaccia I.
\newblock Characterization of strange attractors.
\newblock Physical Review Letters. 1983;50(5):346.

\bibitem{pincus1991approximate}
Pincus SM.
\newblock Approximate entropy as a measure of system complexity.
\newblock Proceedings of the National Academy of Sciences.
  1991;88(6):2297--2301.

\bibitem{bandt2002permutation}
Bandt C, Pompe B.
\newblock Permutation entropy: a natural complexity measure for time series.
\newblock Physical Review Letters. 2002;88(17):174102.

\bibitem{rosso2007distinguishing}
Rosso OA, Larrondo HA, Martin MT, Plastino A, Fuentes MA.
\newblock Distinguishing noise from chaos.
\newblock Physical Review Letters. 2007;99(15):154102.

\bibitem{crutchfield2012between}
Crutchfield JP.
\newblock Between order and chaos.
\newblock Nature Physics. 2012;8(1):17--24.

\bibitem{marcon2014generalization}
Marcon E, Scotti I, H{\'e}rault B, Rossi V, Lang G.
\newblock Generalization of the partitioning of Shannon diversity.
\newblock Plos One. 2014;9(3):e90289.

\bibitem{daw2003review}
Daw CS, Finney CEA, Tracy ER.
\newblock A review of symbolic analysis of experimental data.
\newblock Review of Scientific Instruments. 2003;74(2):915--930.

\bibitem{herzel1994finite}
Herzel H, Schmitt A, Ebeling W.
\newblock Finite sample effects in sequence analysis.
\newblock Chaos, Solitons \& Fractals. 1994;4(1):97--113.

\bibitem{grassberger1988finite}
Grassberger P.
\newblock Finite sample corrections to entropy and dimension estimates.
\newblock Physics Letters A. 1988;128(6-7):369--373.

\bibitem{stammler2016correcting}
Stammler S, Katzenbeisser S, Hamacher K.
\newblock Correcting Finite Sampling Issues in Entropy l-diversity.
\newblock In: International Conference on Privacy in Statistical Databases.
  Springer; 2016. p. 135--146.

\bibitem{levina2017subsampling}
Levina A, Priesemann V.
\newblock Subsampling scaling.
\newblock Nature communications. 2017;8:15140.

\end{thebibliography}

\section*{Supporting information}

\label{S1}
 In this work, six different financial data sets have been investigated:  BUND, BOBL, DAX,   EuroCurrency, EuroSTOXX, FIB 30. A short description of the data and the markets is provided here below.\\
\par

{\bf BUND}  is a debt security issued by Germany's federal government, and it is the German equivalent of a U.S. Treasury bond. The German government uses bunds to finance its spending, and bonds with long-term durations are the most widely issued securities. Bunds are auctioned only with original maturities of 10 and 30 years.
Bunds represent long-term obligations of the German federal government that are auctioned off in the primary market and traded in the secondary market. Bunds can be stripped, meaning their coupon payments can be separated from their principal repayments and traded individually. Bunds pay interest and principal typically once a year, and they represent an important source of financing for the German government.
The principal characteristics of bunds are that they are nominal bonds with fixed maturities and fixed interest rates. All German government debt instruments, including bunds, are issued by making a claim in the government debt register rather than producing paper certificates. A typical bund issue will state its issuance volume, maturity date, coupon rate, payable terms and interest calculation standard used.\\
%
\par
{\bf BOBL} is a futures contract with medium term debt that is issued by the Federal Republic of Germany as its underlying asset. The contract has a notional contract value of 100,000 euros, with a term to maturity of 4.5 to five years. Unlike most other types of future contracts, BOBL future contracts tend to be settled by delivery.
In America, these futures contracts are traded on the Chicago Board of Trade, under the symbol GBM. The Euro-Bobl future, along with the Euro-Bund and Euro-Schatz futures, are the most heavily traded fixed-income securities in the world.
BOBL is an acronym for bundesobligationen. This translates to 'federal obligations' in English .\\
\par
{\bf DAX}, the ``Deutscher Aktien indeX'', is the derivative of the main German stock index consisting of the 30 major German companies trading on the Frankfurt Stock Exchange.
Just like the FTSE 100 and S\&P500, DAX is a capitalization-weighted index so it essentially measures the performance of the 30 largest, publicly traded companies in Germany. It is therefore a strong indicator of the strength of the German economy and investor sentiment towards German equities.
The DAX has been a relatively stable index with 16 companies of the original 30 remaining in the index since its inception in 1988.
The index began with a base date of 30 December 1987 and a base value of 1,000. Over the years the DAX has seen a large amount of takeovers, mergers, bankruptcies, and restructurings.
Since 1 January 2006, the index is calculated after every seconds. It is computed daily between 09:00 and 17:30 CET. \\
\par
{\bf Euro Stoxx} is an index of the Eurozone stocks designed by STOXX, a provider owned by Deutsche Börse Group. The Euro Stoxx 50 provides a blue-chip representation of Supersector leaders in the Eurozone. It is made up of fifty of the largest and most liquid stocks.
The Euro Stoxx 50 was introduced on 26 February 1998. Its composition is reviewed annually in September. The index is available in several currency (EUR, USD, CAD, GBP, JPY) and return (Price, Net Return, Gross Return) variant combinations. Calculation takes place every 15 seconds between 09:00 CET and 18:00 CET for the EUR and USD variants of any return type, while the CAD, GBP and JPY variants are available as end-of-day calculation only (18:00 CET).
The Euro Stoxx 50 Index is derived from the 19 Euro Stoxx regional Supersector indices and represents the largest super-sector leaders in the Eurozone in terms of free-float market capitalization. The index captures about 60\% of the free-float market capitalization of the Euro Stoxx Total Market Index (TMI), which in turn covers about 95\% of the free-float market capitalizaion of the represented countries.
It is one of the most liquid indices for the Eurozone: an ideal underlying for financial products or for benchmarking purposes. Additionally, the index serves as an underlying for many strategy indices, such as the Euro Stoxx 50 Risk Control Indices. Buffers are used to achieve the fixed number of components and to maintain stability of the indices by reducing index composition changes. Selection methodology ensures a stable and up-to-date index composition. Fast-entry and fast-exit rules ensure the index accurately represents the performance of only the biggest and most liquid stocks.\\
\par
{\bf Euro Currency Index}   ($EUR_I$) represents the arithmetic ratio of four major currencies against the Euro: US-Dollar, British Pound, Japanese Yen and Swiss Franc. All currencies are expressed in units of currency per Euro. The index was launched in 2004 by the exchange portal Stooq.com. Underlying are 100 points on 4 January 1971.
Based on the progression, Euro Currency Index can show the strength or weakness of the Euro. A rising index indicates an appreciation of the Euro against the currencies in the currency basket, a falling index in contrast, a devaluation. Relationships to commodity indices are recognizable. A rising Euro Currency Index means a tendency of falling commodity prices. This is especially true for agricultural commodities and the price of oil. Even the prices of precious metals (gold and silver) are correlated with the index.
Arithmetically weighted Euro Currency Index is comparable to the trade-weighted Euro Effective exchange rate index of the European Central Bank (ECB). The index of ECB measures much more accurately the value of the Euro, compared to the Euro Currency Index, since the competitiveness of European goods in comparison to other countries and trading partners is included in it.\\
\par
{\bf FIB30} is the derivative of the MIB 30 stock market index consisting of the 30 major Italian companies trading on the Borsa Italiana until the 2003. In fact in 2003 it is substituted by the S\&P/MIB, and in the 2009 by the FTSE-MIB index based on the 40 major company traded in the Italian market.
The FTSE MIB index is the major benchmark index of the Italian stock markets. This index, which accounts for about 80\% of the domestic market capitalization, consists of companies of primary importance and high liquidity in the various ICB sectors in Italy.
The FTSE MIB Index measures the performance of 40 Italian securities and intends to reproduce the weights of the expanded Italian equity market. The Index is derived from the trading universe of securities on the main stock market of Borsa Italiana (BIt). Each title is analyzed by size and liquidity, and the Index provides a representation of sectors. The FTSE MIB is a weighted index based on market capitalization.
It is computed daily between 9:00 and 17:50 CET.\\

\end{document}